\documentclass{llncs}

\usepackage{cite}

\usepackage{listings}
\usepackage{color}
\definecolor{light-gray}{RGB}{179,179,179}

\usepackage{graphicx}
\graphicspath{{./figures/}}
\usepackage{amssymb} 

\usepackage[llncs,crop]{llncsconf}
\conference{}
\llncs{This is a post-peer-review, pre-copyedit version of an article published in Lecture Notes in Computer Science (LNCS, volume 10437).}{1}
\llncsdoi{10.1007/978-3-319-64119-5_14}

\begin{document}

\lstset{
 language=XML,
 tabsize=2,
 showspaces=false,
 showstringspaces=false,
 basicstyle=\ttfamily\scriptsize,
 float=[htb],
 captionpos=b,
 frame=shadowbox,
 rulesepcolor=\color{light-gray},
 numbers=left,
 numberstyle=\tiny,
 breaklines=true,
 linewidth=0.96\columnwidth,
 xleftmargin=5.0ex
}

\title{Verification of Component Fault Trees using Error Effect Simulations}

\author{Sebastian Reiter\inst{1}, Marc Zeller\inst{2}, Kai H{\"o}fig\inst{2}, Alexander Viehl\inst{1},\\ Oliver Bringmann\inst{1}, Wolfgang Rosenstiel\inst{1}}

\institute{ FZI Forschungszentrum Informatik \\
Haid-und-Neu-Str. 10-14, D-76131 Karlsruhe, Germany \\
\{sreiter, viehl, bringman, rosenstiel\}@fzi.de \and
Siemens AG, Corporate Technology \\
Otto-Hahn-Ring 6, D-81379 Munich, Germany \\
\{marc.zeller,kai.hoefig\}@siemens.com
}

\maketitle

\begin{abstract}
The growing complexity of safety-relevant systems causes an increasing effort for safety assurance.  
The reduction of development costs and time-to-market, while guaranteeing safe operation, is therefore a major challenge.
In order to enable efficient safety assessment of complex architectures, we present an approach, which combines deductive safety analyses, in form of Component Fault Trees (CFTs), with an Error Effect Simulation (EES) for sanity checks.
Both CFTs and the EES provide a modular, reusable and compositional safety analysis. 
The combination reduces the drawbacks of both analyses, such as the subjective failure propagation assumptions in the CFTs or the determination of relevant fault scenarios for the EES. 
Since both are applicable throughout the whole design process, they support continuous model refinement and the reuse of conducted safety analysis and simulation models. 
Hence, safety goal violations can be identified in early design stages and the reuse of conducted safety analyses reduces the overhead for safety assessment. 
\end{abstract}

\section{Introduction}
\label{sec:introduction}
The growing number and complexity of safety-relevant embedded systems poses new challenges, in many application domains such as the automotive domain with the rapidly evolving advance driver assistance systems. 
Along with the growing system complexity, also the need for safety assessment and its associated effort is drastically increasing.
Safety assessment is a mandatory part in order to guarantee the high quality demands, imposed by the market.
However, this is contrary to industry's aim to reduce costs and time-to-market.
\\In different application domains, safety standards, such as the IEC~61508~\cite{iec61508} or its automotive adaption ISO~26262~\cite{iso26262} define the safety assurance process.
The goal of the safety assessment process is to identify all failures that cause hazardous situations and to demonstrate that their probabilities are sufficiently low.
Traditionally, the analysis of systems in terms of safety consists of bottom-up safety analysis approaches, such as \textit{Failure Mode and Effect Analysis (FMEA)}, and top-down ones, such as \textit{Fault Tree Analysis (FTA)}.
Both provide structured procedure models to identify failure modes, failure causes and the effects on the system safety goals.
The early identification of safety goal violations is crucial for a cost-efficient development process.

The quality of the safety assessment strongly depends on the correctness and completeness of the safety analysis model, describing the failure propagation and its transformation through the system. 
In current industrial practice, the quality of the safety analysis models in terms of completeness and correctness needs to be guaranteed manually by model reviews, which are very time-consuming tasks. 
Since such reviews are required after the initial construction of the safety analysis model as well as after each modification, the effort for maintaining its quality during the whole design time is significant. 
Applying a development strategy with frequent changes, such as agile methods, drastically increases this maintenance effort, endangering the success of the development.

Since system-level simulations enable the simultaneous analysis of software and digital/analog hardware, the application of simulations for safety evaluations attracted growing attention \cite{6899147,6927251,6881440,5474177,reiter2013,4341528,4669223}.
However, the focus of most of these approaches is either the verification of fault tolerance mechanisms or the identification of failure modes as well as their effects.
The determination of the analyzed fault space, as well as the integration in the still required safety assessment process is neglected.
The main contribution of this paper is the synergistic combination of both methodologies.
We present an approach for the verification of a failure propagation specification in form of a \emph{Component Fault Tree (CFT)} methodology~\cite{Kaiser2003} with a \emph{Error Effect Simulation (EES)} approach. 
Test cases are automatically generated from CFTs, which serve as input to inject faults into a system simulation.
Additionally a test oracle, for failure effect detection, is created automatically.
Thus, it is possible to discover gaps and errors in the failure propagation specification.
This increases the quality while simultaneously reduces the effort for the manual safety model review drastically.
\\Since system simulations can be performed at different levels of granularity and CFT elements can be reused in different contexts, our approach is applicable throughout the design process. 
Hence, iterative and agile development strategies are supported by speeding up the analysis of the current system design abstraction.
Thereby, only the refined parts of the system must be modified.

The paper is organized as follows:
In Sect.~\ref{sec:relatedwork} relevant related work is briefly summarized.
The Sect.~\ref{sec:methodology} presents the underlying method.
The concept of CFTs is presented in Sect.~\ref{sec:cfts}. 
The EES infrastructure is presented in Sect.~\ref{sec:simulator}. 
The Sect.~\ref{sec:generation} outlines how test cases can be automatically generated from CFTs as input for the EES in order to review the safety analysis model. 
In Sect.~\ref{sec:evaluation} the benefits of our approach are demonstrated.
The paper is concluded in Sect.~\ref{sec:conclusion}.

\section{Related Work}
\label{sec:relatedwork}

As already mentioned, there exist manifold approaches that utilize simulation-based fault injection (SFI), such as \cite{6899147,6927251,6881440,5474177,reiter2013,4341528,4669223}. 
These approaches use system simulations combined with fault injection for enhancing safety analyses.
They focus on the efficient execution of the fault injection and the determination of error effects.
Randomly generated parameters are used in a large number of simulation runs. 
But in order to get meaningful results which can be used as evidence in a safety case, the number of test runs must be extremely high. 
And still these approaches cannot guarantee that all relevant faults are injected into the simulation. 
Additionally, previous approaches are missing a methodology to combine the presented techniques with established safety assessment processes which are still required for system qualification/certification.
SFI mitigates the new challenges, but cannot replace safety assessments such as FTA or FMEA.

Such an FMEA integration is presented in \cite{7167184}.
It presents a simulation-assisted FMEA by formalizing the FMEA structure and deriving a system simulation.
With the help of such an approach, the fault catalogues used during FMEA are reused by the SFI.
We utilize a similar approach for FTA, by mapping the FTA components to simulation components of a SFI.
By provision of both a component-based FTA and a component-based SFI the concept of reusable databases is extended to reusable safety analysis artifacts. 

The formalization of the FTA and the SFI structure are realized with the UML.
Approaches such as \cite{Weissnegger2016, 7027431} are also using the UML to support safety assessment. 
\cite{7027431} provides a model-to-model transformation to map a UML specification to a executable model based on the \emph{Action Language for Foundational UML}.
However, the derivation or construction of FMEA tables are neglected.  

In \cite{Marinescu2015} Simulink-based simulation models are generated based on an EAST-ADL description of the system. 
Safety analysis is explicitly targeted by this approach, but it cannot be used to verify safety analysis. 
Similar to other SFI approaches, it is not described how input for the fault injection is generated.

\section{Methodology}
\label{sec:methodology}
\begin{figure}[t]
  \centering
  \includegraphics[width=.65\textwidth]{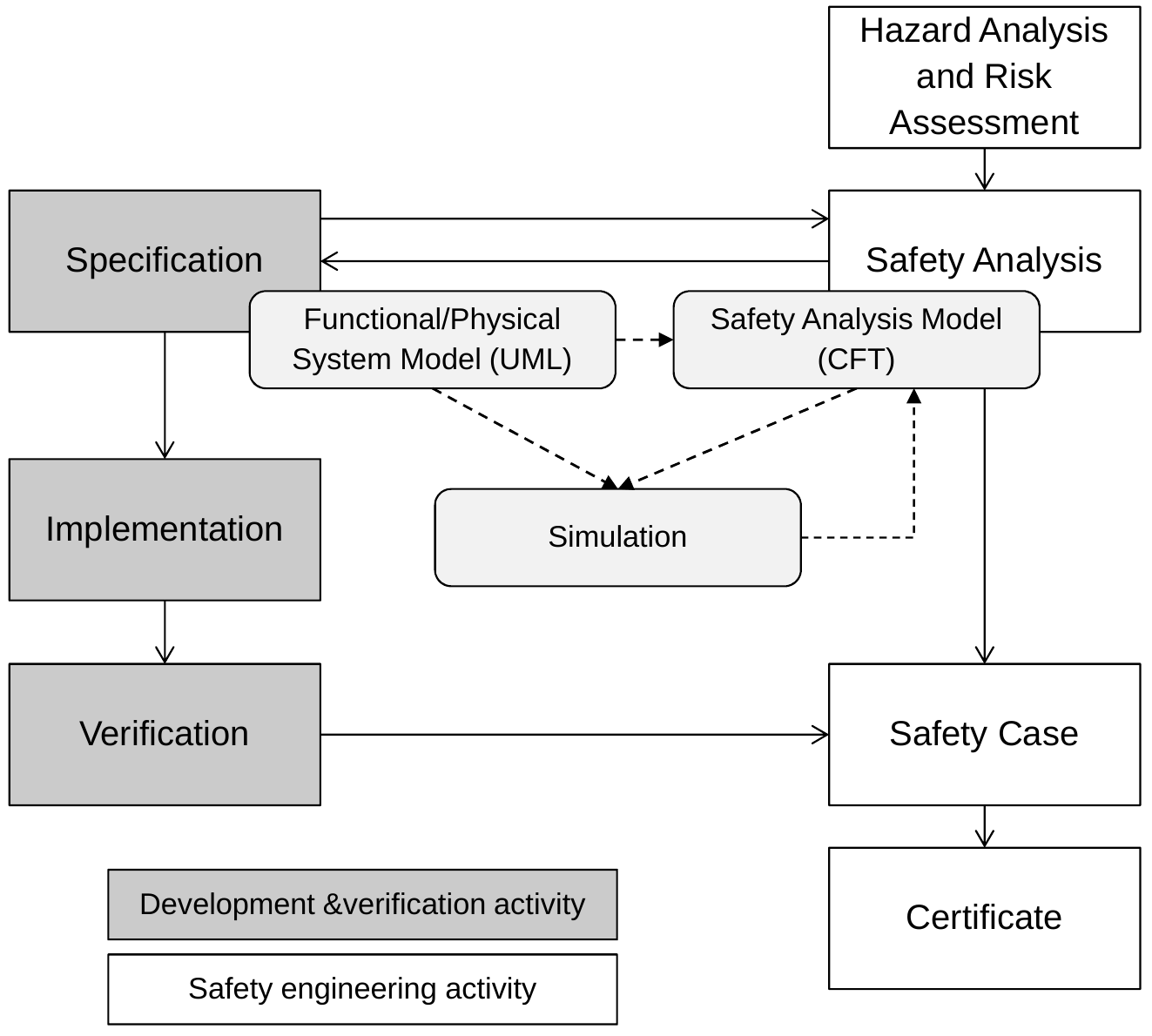}
  \caption{Overview over the proposed methodology to verify safety analysis models}
  \label{fig:methodology}
\end{figure}

The goal of the safety assessment process is to identify all failures that cause hazardous situations and to demonstrate that their probabilities are sufficiently low. In the application domains of safety-relevant systems, the safety assurance process is defined by the means of safety standards (e.g.~IEC~61508~\cite{iec61508}).
Fig.~\ref{fig:methodology} illustrates typical activities and their relation to the development process. 
Moreover, it outlines the used models and their interactions.
As a first step, all relevant hazards of the system are identified and the associated risks are assessed during the so-called \emph{Hazard Analysis and Risk Assessment}. 
Top-level safety requirements are the result of this step. 
Based on the system requirements the architecture of the system is designed in a model-based way using UML. 
Based on this system specification, the safety engineer is developing a safety analysis to identify failure modes, their causes, and effects with impact on the system safety.
The results of the safety analysis as well as results of verification activities are used as evidences to evolve safety requirements to a safety case. 
Finally, the system safety is assessed based on the safety case and a certificate is issued in case of a positive assessment result.

In our approach, we use the CFT methodology as a model- and component-based approach for fault tree analyses to enable modular and compositional safety analysis. 
Thereby, fault tree elements are related to their development artifacts (defined in UML) and can be reused along with them.
Based on the available system specification in UML a system level simulation can be generated automatically.
Since safety artifacts are associated with design artifacts of the system architecture, a straightforward mapping between the entities in the generated system simulation and the safety analysis model is given. 
Hence, test cases with fault injection and test oracles are automatically generated from CFTs. 
This way it possible to check whether the specified failure propagation is correct, using the system-level simulation. 
This enhances the quality of the CFT-based safety analysis model in terms of correctness, since failure modes as well as failure mitigation mechanisms can be discovered which the safety engineer overlooked.
Moreover, failure propagation paths, which are not yet know, e.g.~caused by emergent behavior on system level, can be discovered.

\section{Component Fault Trees}
\label{sec:cfts}
Component Fault Trees are Boolean models associated with system development elements such as components \cite{Kaiser2003,Hoefig2015a,Zeller2016a}. 
It has the same expressive power as classic fault trees, which are described in \cite{vesely1981fault}. 
Like classic fault trees, CFTs are used to model failure behavior of safety-relevant systems. 
This failure behavior, including their appearance rate, is used to document the absence of unreasonable risk of the overall system. 
In addition, it can also be used to identify drawbacks of the design of a system. 
In CFT methodology, a separate \emph{CFT element} is related to a component, e.g. defined in UML \cite{AdlerDHKKST10}. 

Let the System $Sys$ consist of a set of components $C = \left\{ c_{1},...,c_{n} \right\}$. 
Each component $c_i \in C$ has a set of inports $IN(c_i) = \left\{ in_{1},...,in_{p} \right\}$ and a set of outports $OUT(c_i) = \left\{ out_{1},...,out_{q} \right\}$. 
The information flow between the outport of a component $c_i \in C$ and the inport of another component $c_j \in C$ (with $c_i \neq c_j$) is modelled by a set of connections
$CON = \left\{ (out_x, in_y) | out_x \in OUT(c_i) , in_y \in IN(c_j) \right\}$.

If $c_i \in C$ has a component fault tree element $cft_i \in CFT$, then it is $\tilde{CFT}(c_i) = cft_i$ with $cft_i \neq \emptyset$.
Thus, the CFT of the system $Sys$ is defined by the set of CFT elements $CFT = \left\{ cft_1,...,cft_n \right\}$.

Failures that are visible at the outport of a component are modeled using \emph{Output Failure Modes} $OFM(out_l) = \left\{ ofm_1,...,ofm_t \right\}$ which are related to the specific outport $out_l \in OUT(c_i)$. 
To model how specific failures propagate from an inport of a component to the outport, \emph{Input Failure Modes} $IFM(in_k) = \left\{ ifm_1,...,ifm_s \right\}$ are used, which are related to an inport $in_k \in IN(c_i)$. 
The internal failure behavior that also influences the output failure modes is modeled using Boolean gates such as \emph{OR} and \emph{AND} as well as \emph{Basic Events}.
Basic Events $B(cft_i) = \left\{ b_1,...,b_r \right\}$ represent failure modes that originate within a component.
Each Basic Event can be assigned a failure rate, e.g. the \emph{Mean Time Between Failure (MTBF)} or the \emph{Failure in Time (FIT)}.
In case of an OR gate a failure propagates if at least one of the inputs is active, while an AND gate propagates failures only if all input failures are active.

A library, which contains CFT elements for all system components, eases the reusability of safety artifacts. Hence, it is possible to create different CFTs by just changing the assembly of the CFT elements.
Every CFT can be transformed to a classic fault tree by removing the input and output failure mode elements. 

\section{Error Effect Simulation}
\label{sec:simulator}
The EES consists of a \emph{Device Under Test (DUT)} and its test bench. 
The DUT models software, hardware and analog system parts.
We use the IEEE standard \emph{SystemC}~\cite{STD:SystemC} because of its support of various simulation domains and its comprehensive support of different abstraction levels.  
The EES uses an approach similar to \cite{7748256}. 
It consists of a modular, parameterizable system simulation that is used to assess error effects. 
One simulation instance is assembled of parameterizable \emph{Simulation Entities (SEs)}.
To facilitate the analysis of different DUT characteristics, the framework uses dynamic configuration that is supplied during runtime. 
This covers the DUT architecture as well as the SEs parameterization. 

The EES requires \emph{fault injectors}, which cause a discrepancy within the DUT. 
We provide injection capability by replacing simulation primitives with an injectable container that encapsulates the original simulation primitive. 
This container provides an interface to change the value of the simulation primitive.
The design of the injectors is particularly suited to support simulation models of different abstraction levels with their different modeling primitives. 
A centralized injection control module stimulates the safety case by controlling the injectors. 

The so-called \emph{Behavioral Threat Models (BTM)} specifies the injected behavior.
A BTM is based on \emph{Timed Automata (TA)}.
A TA consists of a finite set of locations $L$, actions $\Sigma$ and a set of clocks $C$, which can be reset individually.
An edge $(l_{i},\sigma,g(C),r(C),l_{j} ) \in E$ is a transition from location $l_{i}$ to location $l_{j}$ with an action $\sigma \in \Sigma$, a time dependent guard $g(C)$ and a clock reset $r(C)$.
For fault modeling the TA approach is extended in a way that guard statement $g(C,\mathbb{W}_{s}, \mathbb{W}_{l}, \mathbb{E})$ can additionally depend on the state of the simulation $\mathbb{W}_{s}$, the state of BTM local variables $\mathbb{W}_{l}$ and a set of events $\mathbb{E}$ that are used to synchronize multiple BTMs.  
Actions $\sigma(\mathbb{W}_{s},\mathbb{W}_{l})$ are extended to modify the simulation state $\mathbb{W}_{s}$ or local variables within the BTM $\mathbb{W}_{l}$.
The centralized injection control module interprets a BTM dynamically and stimulates faults.  

For failure detection, the virtual prototype is extended with \emph{failure monitors}, which compare the behavior of one variable with a previous simulation run. 
Using an error free simulation run as reference enables the detection of failures that are caused by the fault injection.
Ref. \cite{tax:avi} presents a general failure mode characterization, which cover \emph{content failures}, \emph{early failures}, \emph{late failures}, \emph{halt failures} as well as \emph{erratic failures}.
The proposed failure monitors detect these standard failure modes by default. 
Besides the detection of these standard failure modes it is possible to specify queries that are checked both on the reference and error trace. 
A \emph{Computation Tree Logic (CTL)} expression specifies the queries. 
This enables the detection of application specific failure modes.  

A graphical specification of the EES based on the UML supports the user by conducting analyses. 
It reduces the manual overhead by code generation steps and increase the usability by a graphical user interface.   
The graphical specification consists of the specification of the SEs, their instantiation, parametrization and interconnection. 
Moreover, the placement of fault injectors, failure monitors and fault behavior is specified.

With these extensions of the system simulation the question still remains, at which location to inject faults (fault injector placement), how does the injected behavior look like (BTM specification) and where to detect deviations from the reference trace (monitor placement).

\section{Verification of CFTs using Error Effect Simulation}
\label{sec:generation}
In order to verify the failure behavior specified as CFT, test cases for the EES are generated automatically. 
A test case covers the fault specification and the fault injector placement. 
Additionally it includes a test oracle, which is realized by a failure monitor. 
The test cases are derived from the CFT.
For this purposes, we define a so-called \emph{scope S} of the CFT that involves only a certain amount of the components with $S \subseteq CFT$.  
Thus, it is possible to verify both the entire failure propagation model as well as only a part of it.
A scope $S \subseteq CFT$ provides a set of inputs and outputs. 
The inputs of the scope $IN_S \subseteq \bigcup_{c_i \in S} IN(c_i)$ are used to enter a test scenario. 
The outputs $OUT_S \subseteq \bigcup_{c_i \in S} OUT(c_i)$ are used to measure the results of a test scenario.  
The inner CFT logic can be simplified to a CFT element for the scope $S$, which only contains the gates and basic events, input and output failure modes that are related to the scope (cf.~Ref.~\cite{Zeller2015a}). 
Internal input and output failure modes are omitted.
Hence, for a scope $S \subseteq CFT$, the CFT element related to $S$ is $CFT_S \subseteq CFT$. 
It contains failure modes related to the inports and outports and have a connection outside of the scope. 
Let
\begin{eqnarray*}
IFM(S)=\{in \; &|& \; \exists (a,b) \in CON, a \in OUT(A), A \notin S, \\
  &&b \in IN(B), B \in S, in \in IFM(B)\}
\end{eqnarray*}
be the input failure modes of the scope $S$ and
\begin{eqnarray*}
OFM(S)=\{out \; &|& \; \exists (a,b) \in CON, a \in OUT(A), A \in S,\\
  &&b \in IN(B), B \notin S, ut \in OFM(A)\}
\end{eqnarray*}
be the output failure modes of the scope $S$.
Moreover, let
\begin{displaymath}
B(S) =   \bigcup_{c_i \in S} B(c_i)
\end{displaymath}
be the set of basic events of the scope $S$.

In order to determine a set of test cases, we apply the methodology of \emph{Minimal Cut Set Analysis (MCA)}. 
A MCA is a representation of a tree using a disjunction of conjunctive terms that cannot be reduced further \cite{vesely1981fault}.
For a scope $S$, let 
\begin{displaymath}
mc_i(t)=x_1\wedge \dots \wedge x_n, t \in OFM(S), x_i \in IFM(S) \cup B(S)
\end{displaymath}
be all cut sets that result in the occurrence of the output failure mode $t \in OFM(S)$ of the scope $S \subseteq CFT$.
Moreover, let 
\begin{displaymath}
MCA(t) = mc_1(t)\vee \dots \vee  mc_m(t), t \in OFM(S)
\end{displaymath}
be the minimal cut set analysis of the output failure mode $t$ and scope $S \subseteq CFT$.
Since in general multiple combinations of input data leads to different output data for the same test case, typical measures can be applied here to further reduce the set of test cases like equivalence class testing.

Thus, the results of the MCA are used to generate input for the fault injectors.
For each input failure mode $ifm_{i}$ a set of BTMs is generated $BTM(ifm_{i}) = \left\{ btm_{i,1} \ldots btm_{i,m} \right\}$ to control the fault injector at the inports $IN_S$ of the respective component. 
Similar for each Basic Events $b_{i}$ a set of BTMs is generated $BTM(b_{i}) = \left\{ btm_{i,1} \ldots btm_{i,n} \right\}$ to inject faults by the modification of the component state. 
In this work, we use sets of predefined BTMs that match the failure behavior of the input failure modes and Basic Events within the CFT. 
An approach with a BTM-library is used and the generation process parametrizes these predefined fault behaviors. 
The generated inputs stimulate the DUT and will lead to a discrepancy from the intended behavior.
Since the output failure modes $OFM(S)$ of $S$ can be observed at the outports $OUT_S$, failure monitors are placed within the EES. 
These monitors detect general failure modes by default but can also be extended to detect application specific failure modes. 
With this the the CFT can be verified by the EES.

\section{Case study}
\label{sec:evaluation}
We illustrate the benefits of our approach using an automotive case study namely a coasting assistant. 
This assistant calculates driver hints such as acceleration or braking with regard to an energy efficient driving strategy.
The system is designed for electric vehicles, where in case of a deceleration the driver has the choice of freewheeling or braking with recuperation. 
The coasting assistant displays the optimal driving choice w.r.t.~energy efficiency in the Head-Unit. 
Therefore, the coasting assistant ECU uses maps annotated with speed limits but also a video-based speed limit detection to adjust to undocumented limits such as road works.
This speed limit detection will be the focus of the case study. 

\subsection{Overview}
\begin{figure}[b]
  \centering
  \includegraphics[width=.6\textwidth]{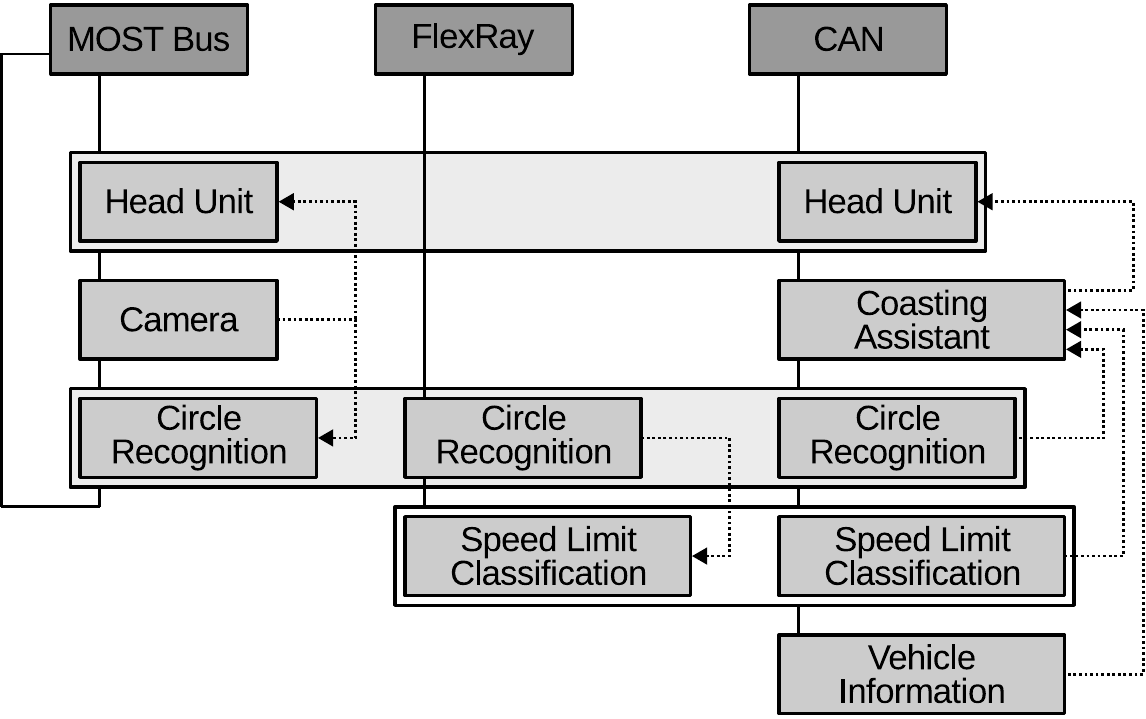}
  \caption{Overview of the case study}
  \label{fig:usecase}
\end{figure}

The coasting assistant functionality is distributed over multiple Electronic Control Units (ECUs) which are interconnected by different communication channels.    
Fig.~\ref{fig:usecase} shows the involved ECUs and communication channels. 
The camera ECU records the road in front of the vehicle and sends this image to a circle detection algorithm using the synchronous channel of a MOST bus.
In the next processing step circles are detected and the containing image segments are forwarded to a speed limit classification. 
The image segments are forwarded via the dynamic segment of the FlexRay bus.
Information about the estimated distance of the potential speed sign is sent with a separate CAN message to the coasting assistant ECU. 
The cropped circles are classified and the results are sent to the coasting assistant via CAN. 
The coasting assistant uses the received information on the next speed limit as well as the estimated distance and an internal look-up table to calculate an energy efficient driving strategy. 

\begin{figure}
  \centering
  \includegraphics[width=.98\textwidth]{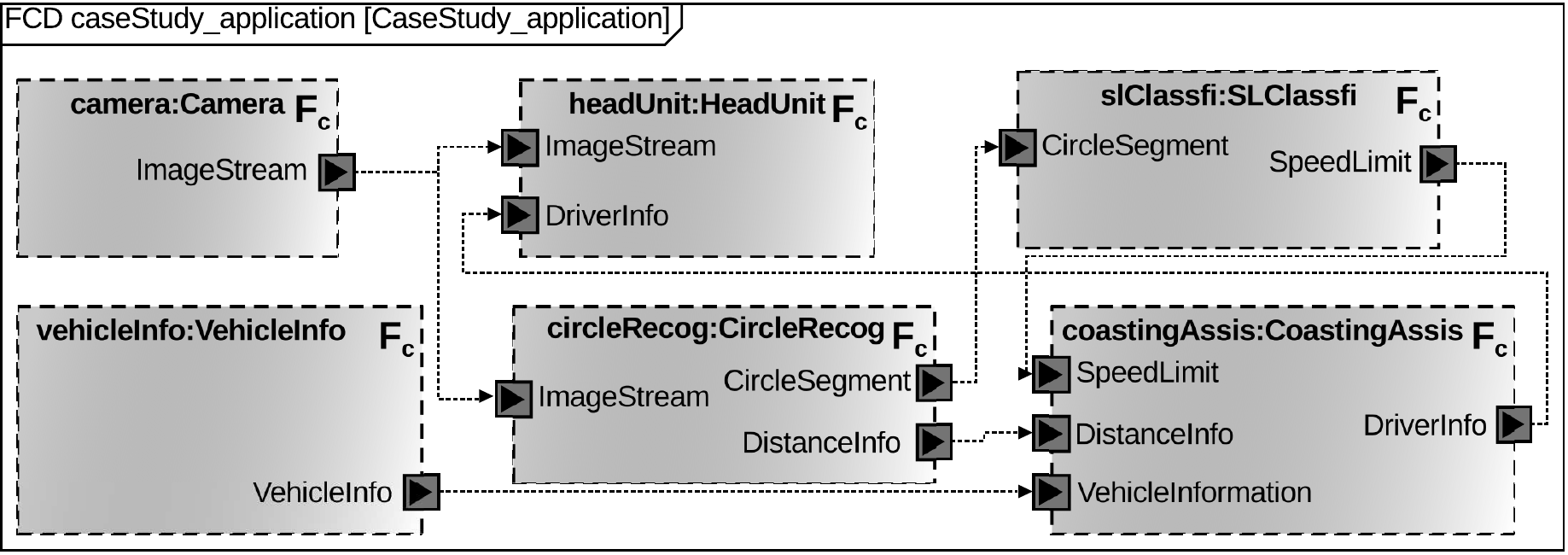}
  \caption{Functional architecture of the exemplary system}
  \label{fig:appCFT}
\end{figure}

Focus of the safety analysis in early development stages is the actual functionality itself.
The communication between simulation entities is abstracted by function calls. 
The logical system architecture, specified with the CFT architectural model, is shown in Fig.~\ref{fig:appCFT}.
The data exchange is modeled using ports.

\subsection{CFT-based analysis}
\begin{figure}[b]
  \centering
  \includegraphics[width=\textwidth]{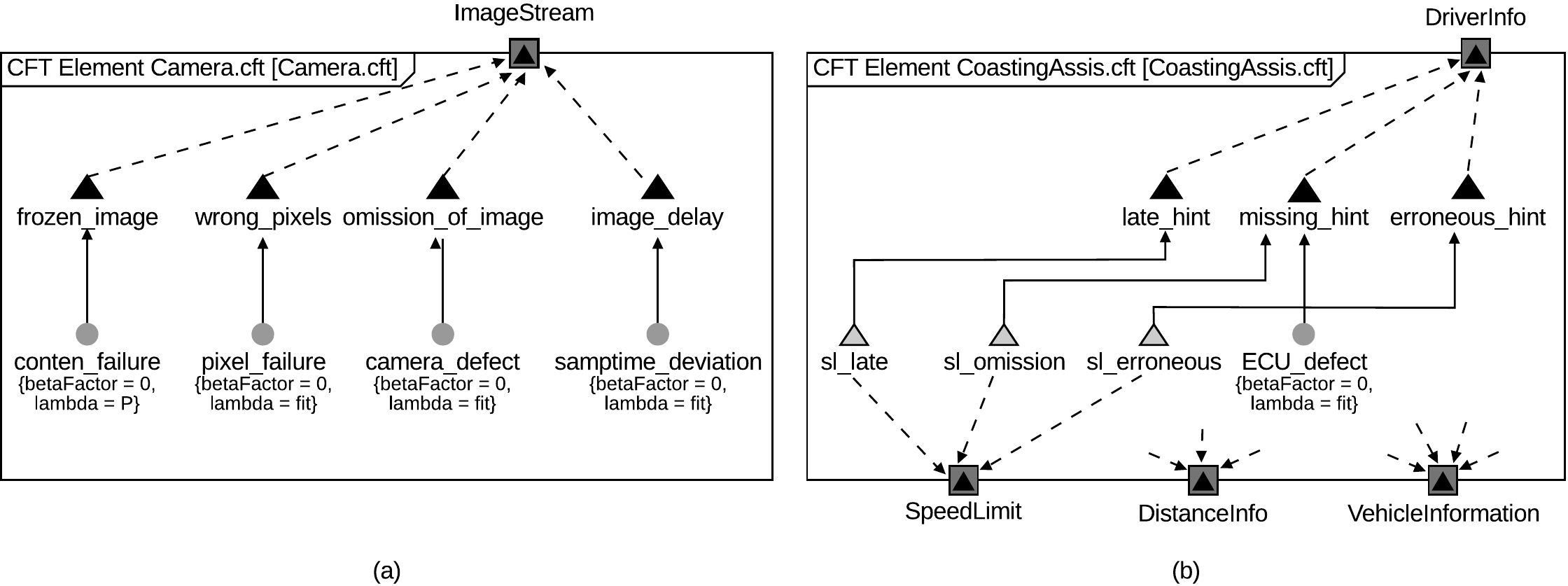}
  \caption{Excerpt of failure modes within the camera (a) and coasting assistant (b)}
  \label{fig:cameraCFT}
\end{figure}

For each of the components within the system architecture a CFT element is specified. 
Fig.~\ref{fig:cameraCFT} (a) shows the CFT element describing the failure behavior of the camera ECU. The different internal failures are modeled by Basic Events. Each of them is associated with a specific failure rate. 
The CFT element of the camera specifies failure modes like a frozen image stream, a corruption of the image pixels, the omission of the complete images or a variation in the sampling time of the camera. These failure modes are propagated via the components' outports to other components.
This way the failures originating from the camera induces the failure modes in the head unit and the Circle Detection ECU. 
Fig.~\ref{fig:cameraCFT} (b) shows the failure modes within the coasting assistant ECU. 
The input failure modes represent that a speed limit is reported too late that an erroneous speed limit is reported or that speed limit is missing.
The output failure modes depend on the input failure modes and the Basic Events of the component. 
The output failure modes determine the input failure modes of the next component - in our case study the ones of the head unit.

With the specification of the CFT elements for all components of the system architecture it is possible to calculate the complete fault tree or different fault trees w.r.t.~a restricted scope. 
Table~\ref{tab:genTestCases} shows a few of the generated test cases with different scopes.
\begin{table}
\centering
\begin{tabular}{|p{0.2\textwidth}|p{0.4\textwidth}|p{0.33\textwidth}|} \hline
Scope                & Fault injection                                   & Failure monitor                  \\ \hline \hline
camera               & camera\_defect                                    & omission\_of\_image              \\ \hline 
                     & content\_failure                                  & frozen\_image                    \\ \hline
coastingAssistant    & sl\_omission OR di\_omission OR ECU\_defect       & missing\_hint                    \\ \hline
                     & sl\_errornous                                     & erroneous\_hint                  \\ \hline
camera, circleRecog, slClassfi, coastingAssis 
                     & camera.content\_failure OR camera.pixel\_failure  & coastingAssis.erroneous\_hint    \\ \hline                
complete System      & camera.content\_failure OR camera.pixel\_failure  & HMI.image\_content\_failure      \\ \hline   
                     & camera.samptime\_deviation                        & HMI.sl\_late                     \\ \hline

\end{tabular}
\caption{Excerpt of the generated test cases}
\label{tab:genTestCases}
\vspace*{-5mm}
\end{table}
One example fault tree path would be that \texttt{camera.content\_failure OR camera.pixel\_failure} can cause the failure mode \texttt{coastingAssis.erroneous\_hint}, when the HMI is excluded from the analyzed scope. 

\subsection{Simulation-based verification}
\begin{figure}[t]
  \centering
  \includegraphics[width=.7\textwidth]{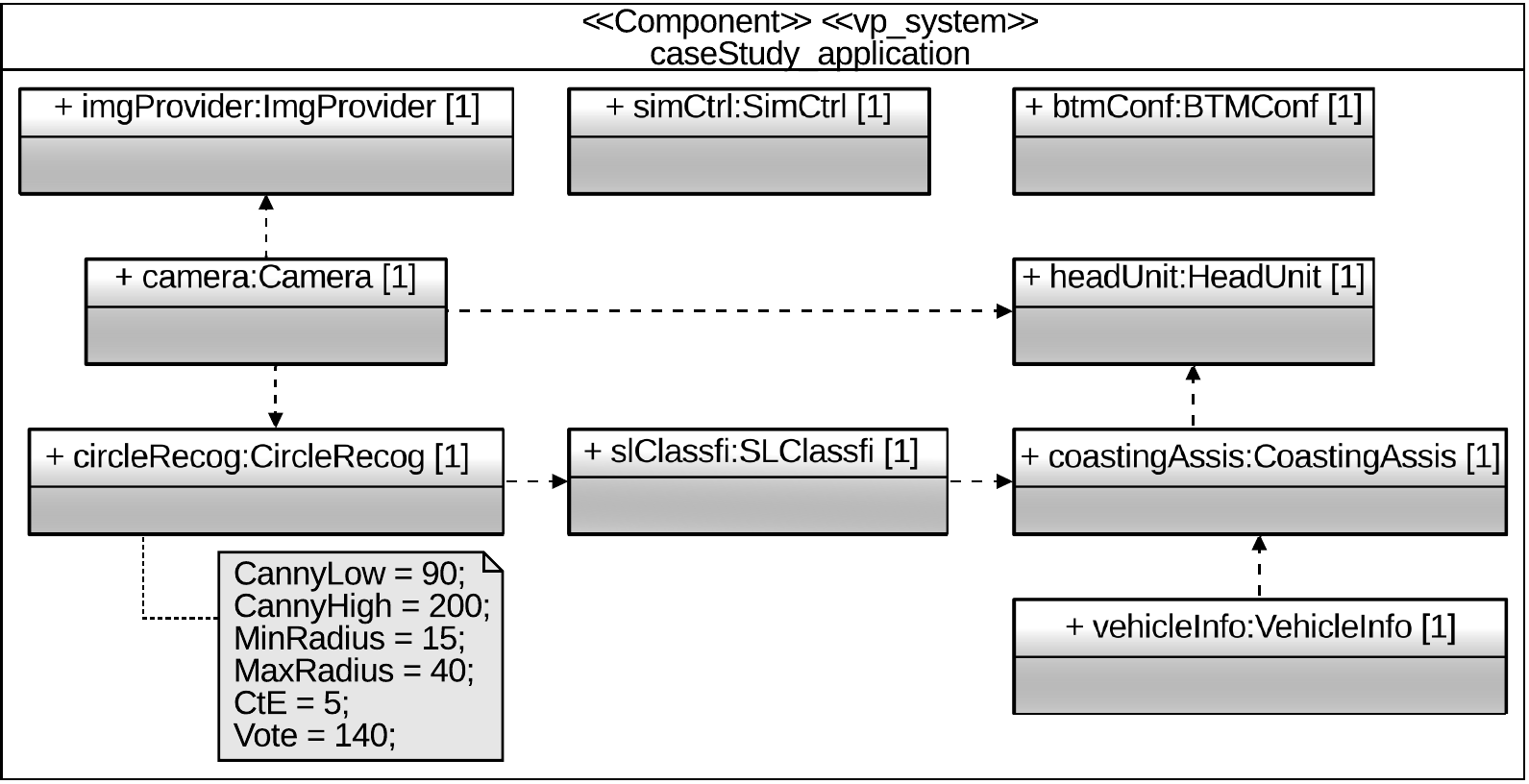}
  \caption{Specification of the system simulation for the case study}
  \label{fig:fes_componentDia}
\end{figure}

After generation of the relation between input failure modes and output failure modes, these relations are verified using simulations. 
The EES shows if the caused-by-relation specified in the safety analysis model is observed in the simulation.  
In case of OR-relations the simulation has to provoke an error effect if any of the faults is injected. 
In case of an AND-relation the error effect should only manifest if all input faults are injected simultaneous. 
For each component in the CFT a corresponding SE exists in the simulation.
Fig.~\ref{fig:fes_componentDia} shows the graphical specification of the simulation configuration. 
It contains the same components as the CFT system architecture and for each of the components a CFT element is specified. 
Model-to-model transformations enable the mapping of CFT elements to simulation components. 
The simulation specification contains additional information that is not specified within the CFT system architecture or the CFT, such as the SE parametrizations or the SEs of the test bench. 
Thus, it is no bijective transformation.
In the context of this work the developer is supported by automated model-to-model transformations, but still manual extensions, e.g. to specify the component parametrization or to extend the configuration with sole simulation components, are required to create the simulation. 
This approach offers the benefit that the system architecture or the CFT is not polluted with information which is only required for the simulation. 

\begin{figure}[t]
  \centering
  \includegraphics[width=.7\textwidth]{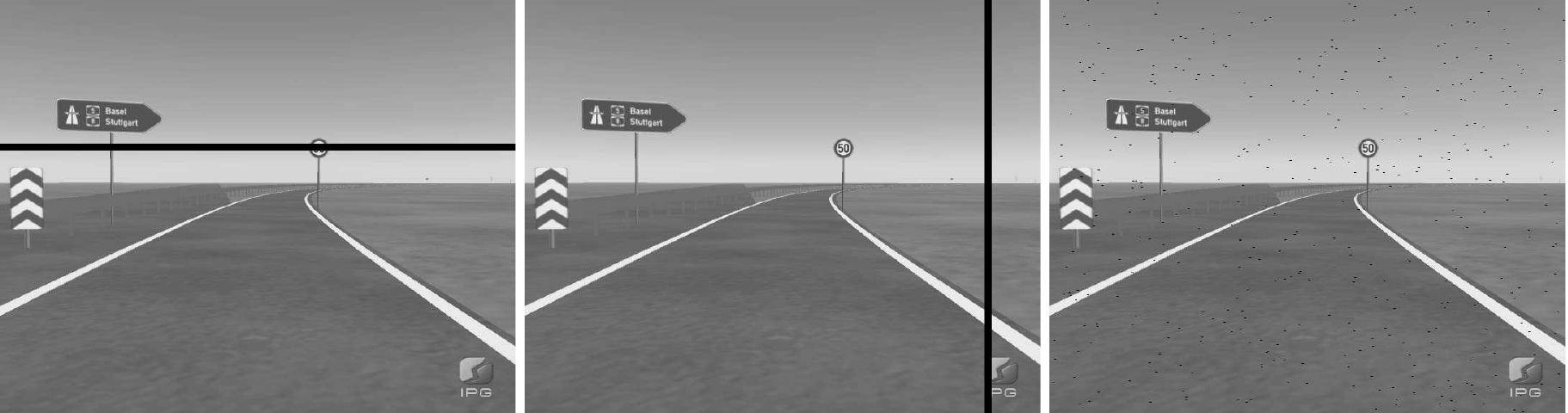}
  \caption{Effect of different pixel corruption strategies}
  \label{fig:bitErrorTrace}
\end{figure}

The fault injection targets mainly the internal information of the components.
The failure modes specified in the CFT element of the Camera
(cf.~Fig.~\ref{fig:cameraCFT}) target the internal storage of the transmitted image.
The failure mode \texttt{frozen\_image} is represented by altering the sampling time of the camera, the \texttt{wrong\_pixels} fault by altering the image payload. 
A camera defect (\texttt{omission\_of\_image}) is modeled by an infinite sampling time. 
If two failure modes are affecting the same injector, e.g. the camera defect and the frozen image, the user have to assure the interaction between the two failures modes is desired.
The simulation contains therefore fault injectors that are associated with the payload array and with the sampling time variable.
Fig.~\ref{fig:btm_pixelFault} (above) shows the BTMs to inject a pixel failure and Fig.~\ref{fig:btm_pixelFault} (below) shows the behavior of the content failure.
\begin{figure}
\begin{lstlisting}[language=XML]
<ESname> eInit <EStype> initial 
<ESname> eFree 
<ESname> eState 
<ETsrc> eInit <ETtgt> eFree     
  <ETaction> ids = [id for id in range(719*200,719*210)] 
<ETsrc> eFree <ETtgt> eState     
  <ETguard> BTMclock('OKTime').read()==sc_time(23, sc_time_unit.SC_SEC)    
  <ETaction> [VPvar('m_Camera.pixel')[id].force(0x00) for id in ids] 
<ETsrc> eState <ETtgt> eState     
  <ETaction> [VPvar('m_Camera.pixel')[id].release() for id in ids];                       
             [VPvar('m_Camera.pixel')[id].force(0x00) for id in ids]
\end{lstlisting}
\begin{lstlisting}[language=XML]
<ESname>errFree <EStype>initial
<ESname>errState1
<ETsrc>errFree <ETtgt>errState1
  <ETguard> BTMclock('OKTime').read()==sc_time(22,sc_time_unit.SC_SEC)
  <ETaction> VPvar('m_Camera.pixel')[0].force(
              VPvar('m_Camera.pixel')[0].read())
\end{lstlisting}
  \caption{BTM for image corruption (above) and a freeze frame failure (below)}
  \label{fig:btm_pixelFault}
\end{figure}
It can be seen that both BTMs control the injector \texttt{VPvar('m\_Camera.pixel')}.  
In the context of the CFT the different failure modes determine the selection of the BTMs ($BTM(ifm_{i})$ or $BTM(b{i})$). 
In this case study, for each input failure mode or Basic Event in the CFT a set of BTM are created.
For instance, the BTM in Fig.~\ref{fig:btm_pixelFault} (below) is generated for the \texttt{camera.content\_failure} solely.
For the \texttt{camera.pixel\_failure} different BTMs that stimulate different corruption patterns are created. 
Fig.~\ref{fig:bitErrorTrace} illustrates three effects of the generated BTMs. 
The picture on the left shows the pixel corruption limited to a horizontal line, the middle one to a vertical line and the picture on the right shows a random pixel corruption over the complete image. 

Each output failure mode within the CFT is presented in the simulation by a monitor.
In this case study a monitor is added at the received speed limit in the coasting assistant, corresponding to the output failure mode specified in the CFT element of the coasting assistant. 
Before each simulation with fault injection a reference trace is generated without fault injection. 
Based on this generated reference trace the monitors classify the current trace, caused by the fault injection.

In this example the traces indicate a late failure. 
The reference trace reports a speed limit at $28.8$ seconds but the injection trace first at $29.1$ seconds.
This is based on the fact that the speed sign is recorded in sequence of images, while the vehicle drives towards the sign, and that the first classifications is prevented based on the pixel corruption within the image. 
In this fault scenario the horizontal line is injected, blocking the speed limits far away. 
This situation was overseen during the creation of the CFT. 
As mentioned earlier a \texttt{camera.content\_failure OR camera.pixel\_failure} can cause the failure mode \texttt{coastingAssis.erroneous\_hint}.
The simulation has shown that this is a very unlikely case, because the classification will most likely return no classification at all instead of the wrong classification. 
Resulting in a late failure mode because a sequence of images is taken from the actual speed limit sign and most likely a subsequent image will be classified correctly.
The nearer the speed limit the more robust is the classification against pixel errors.  
Subsequent to the EES the CFT, particularly the classification component, is adjusted in a way that the \texttt{classification.erroneous\_circle\_recognition} input failure additionally causes \texttt{classification.speed-limit\_information-too-late} output failures. 
This case is a very good example how the none-complete simulation approach and the deductive safety analysis in form of a CFT complement each other. 
A very large number of simulation runs would be necessary to reproduce the behavior of corrupted speed limit information within the coasting assistant based on a pixel corruption in the camera.  
Therefore, the safety analyst can specify this case manually in the fault tree after verifying the CFT using an EES. 
On the other hand when solely using the top-down safety analysis (in form of a CFT), the safety experts could have missed the propagation of a pixel corruption to a late speed limit failure mode.

Besides the verification of existing failure propagation paths within the CFT it is also possible to verify if failure propagation exists that are not represented in the CFT. 
Therefore, the input failure modes and Basic Events, respectively the generated fault behavior, are combined with different output failure modes, respectively their failure monitors. 
Thus, missing propagation paths within the CFT can be revealed using the EES.
With increasing complexity of safety analysis models the effort to verify all combinations of fault injection and monitors is rapidly increasing. 
One benefit of our approach is that different failure monitors can be used within one simulation run in order to examine different failure propagation paths at once.

\subsection{Analyses refinement}
In later stages of the design process the models will be refined. 
In this case study only the communication between the components is refined. 
In a first refinement step, the communication is mapped to abstract transaction models and in a second step to detailed bus models of CAN, FlexRay and MOST.
With each more detailed abstraction level, new failure sources are added and the CFT is extended with CFT elements of the corresponding components. 
This will enable more detailed safety analyses of the system.  
On the other hand the already verified CFTs and simulation models of the application are reused with each refinement step.
This is an example of the benefit of using a modular, compositional approach for the safety analysis and the simulation. 
\\Another benefit is the reuse of components (CFT, EES) within a single system configuration. 
For example the models of the transaction-based communication are required at different locations of the system. 
Through a single CFT- and simulation-element it is possible to create multiple instances.
This way the effort is reduced and the failures are automatically applied at different system parts.

\section{Summary}
\label{sec:conclusion}
The presented approach enables the verification of failure propagation models in form of Component Fault Trees (CFT) by an Error Effect Simulations (EES). 
Automatic generation steps are provided to generate the injected fault behavior, required for the EES, as well as a test oracle to classify the monitored failure behavior. 
It is shown how the verification of the safety analysis model is applied during different phases of the development process. 
In particular, after each change of the system design, including modifications or refinements of the system architecture.  
Since the inputs of the fault injection as well as the test oracle are generated automatically, the repeated verification of the failure propagation model can be performed efficiently.
\\Supporting a safety analysis technique, like CFT, with a non-complete simulation-based analysis reduces the weaknesses of both methodologies. 
The results of the CFT-based analysis are only as good as the system knowledge of the involved experts. 
Especially, the safety analysis model of large-scale, complex systems might be incomplete. 
The system simulation on the other hand offers a method to enable different stakeholders, in particular IP providers, to contribute their system knowledge. 
Such simulation models offer a good basis to support safety analysis, since they enable a quantitative assessment of failure mitigation mechanisms. 
On the other hand, only selective use cases are analyzed and the detailed assessment of failure rates can cause a huge simulation effort.
By the automatic generation of fault injection input and test oracles for the simulation from the CFT these weaknesses are reduced.
Hence, our approach enables the verification whether the specified failure propagation is defined correctly within a CFT-based safety analysis model. 
Moreover, additional failures on system level as well as additional failure mitigation mechanisms can be discovered. 
It is also possible to discover failure propagation paths, which are not yet know.
Another benefit is that in most cases the simulation doesn't have to be executed with a random fault injection based on very small failure rates.  
Firm fault combinations are injected and the propagation is evaluated by the EES. 
The probabilistic analyses is then executed on the verified CFTs, reducing the simulation effort drastically. 

\bibliographystyle{plain}
\bibliography{cft_fes}

\end{document}